\documentclass[
reprint,           
superscriptaddress,
amsmath,           
amssymb,           
aps,               
prd,               
notitlepage,       
floatfix,          
nofootinbib 
]{revtex4-1}

\usepackage{tensor}     
\usepackage{graphicx}   
\usepackage[
colorlinks=true,        
citecolor=blue,         
linkcolor=blue,         
urlcolor=blue           
]{hyperref}             
\usepackage{bm}         
\usepackage{xcolor}     
\usepackage{tabulary}
\usepackage{color}      
\usepackage[utf8]{inputenc} 
\usepackage[section]{placeins} 
\usepackage{ulem}
\usepackage{makecell}
\usepackage{multirow}
\newcommand{\nc}{\newcommand*} 

\nc{\al}{\alpha}
\nc{\s}{\sigma}
\nc{\dt}{\delta}
\nc{\Dt}{\Delta}
\nc{\Ld}{\Lambda}
\nc{\p}{\partial}
\nc{\om}{\omega}
\nc{\Om}{\Omega}
\nc{\rd}{\mathrm{d}}
\nc{\Od}[1]{\mathcal{O}(#1)} 
\nc{\kp}{\kappa}
\nc{\one}{\uppercase\expandafter{\romannumeral1}}
\nc{\two}{\uppercase\expandafter{\romannumeral2}}
\nc{\three}{\uppercase\expandafter{\romannumeral3}}
\def\({\left(}
\def\){\right)}
\def\[{\left[}
\def\]{\right]}
\def\e{\begin{equation}}
\def\q{\end{equation}}
\def\m{\begin{eqnarray}}
\def\n{\end{eqnarray}}
\nc{\Eq}[1]{Eq.~\eqref{#1}}     
\nc{\Fig}[1]{Fig.~\ref{#1}}     
\nc{\Table}[1]{Table~\ref{#1}}  
\nc{\Sec}[1]{Sec.~\ref{#1}}     
\nc{\Msun}{M_\odot}             
\nc{\fpbh}{f_{\mathrm{pbh}}}    
\nc{\fpbhn}{f_{\mathrm{pbh0}}}    
\nc{\mR}{\mathcal{R}} 
\nc{\seq}{\sigma_{\mathrm{eq}}}
\nc{\ogw}{\Omega_{\mathrm{GW}}}
\nc{\gpcyr}{\mathrm{Gpc}^{-3}\,\mathrm{yr}^{-1}}
\nc{\lvc}{LIGO/Virgo} 
\nc{\SNR}{\mathrm{SNR}} 
\nc{\mmin}{{m_{\mathrm{min}}}}
\nc{\mmax}{{m_{\mathrm{max}}}}
\nc{\Mmin}{{M_{\mathrm{min}}}}
\nc{\fmin}{{f_{\mathrm{min}}}}
\nc{\VT}{\mathrm{VT}}
\nc{\rhoGW}{\rho_{\mathrm{GW}}}
\nc{\vth}{\vec{\theta}}
\nc{\vd}{\vec{d}}
\nc{\vla}{\vec{\lambda}}
\nc{\Nobs}{N_{\mathrm{obs}}}
\nc{\av}[1]{\langle #1 \rangle} 
\nc{\km}{\mathrm{km}}
\nc{\Mpc}{\mathrm{Mpc}}
\nc{\Tobs}{T_{\mathrm{obs}}}
\nc{\Ntemp}{N_{\mathrm{temp}}}
\nc{\addref}{[\textcolor{red}{add ref}] } 
\nc{\eg}{\textit{e.g.~}}
\nc{\app}{\approx}
\nc{\hf}{\frac{1}{2}}
\nc{\discuss}{\textcolor{red}{Add discussion here!}}
\nc{\red}[1]{\textcolor{red}{#1}}
\nc{\mH}{\mathcal{H}}
\nc{\cs}{c_s^2}
\nc{\Sij}[1]{S_{ij}^{(#1)}}
\nc{\vi}[1]{v_i^{(#1)}}
\nc{\no}{\nonumber}
\def\<{\left\langle}
\def\>{\right\rangle}

\nc{\bk}{\bm{k}}
\nc{\bq}{\bm{q}}
\nc{\bp}{\bm{p}}
\nc{\bl}{\bm{l}}
\nc{\bx}{\bm{x}}
\nc{\be}{\mathbf{e}}
\nc{\mS}{\mathcal{S}}
\nc{\te}{\tilde{\eta}}
\nc{\tp}{\tilde{p}}
\nc{\tk}{\tilde{k}}
\nc{\tx}{\tilde{x}}
\nc{\tF}{\tilde{F}}
\nc{\tA}{\tilde{A}}
\nc{\mkpq}{|\bk-\bp-\bq|}
\nc{\mpq}{|\bp-\bq|}
\nc{\mkp}{|\bk-\bp|}
\nc{\mSi}[1]{\mS^{(#1)}({\bk, \eta})}
\nc{\vk}{\vec{k}}
\nc{\kstar}{k_*}
\nc{\xstar}{x_*}
\nc{\mpbh}{m_{\rm{pbh}}}
\nc{\Ci}{\mathrm{Ci}}
\nc{\Si}{\mathrm{Si}}
\nc{\fnl}{f_\mathrm{NL}}
\nc{\gnl}{g_\mathrm{NL}}
\nc{\Fnl}{F_\mathrm{NL}}
\nc{\Gnl}{G_\mathrm{NL}}
\nc{\togw}{\tilde{\Omega}}
\nc{\md}{\mathrm{d}^3}
\nc{\taugw}{\tau_{\mathrm{GW}}}
\nc{\tauinst}{\tau_{\mathrm{inst}}}
\nc{\MSmax}{M_{S}^{\mathrm{max}}}

\renewcommand{\vec}[1]{\boldsymbol{#1}} 

\begin{document}
	

\title{Constraints on the ultralight scalar boson from Advanced LIGO and Advanced Virgo's first three observing runs using the stochastic gravitational-wave background}

\author{Chen Yuan}
\email{yuanchen@itp.ac.cn}
\affiliation{CAS Key Laboratory of Theoretical Physics, 
	Institute of Theoretical Physics, Chinese Academy of Sciences,
	Beijing 100190, China}
\affiliation{School of Physical Sciences, 
	University of Chinese Academy of Sciences, 
	No. 19A Yuquan Road, Beijing 100049, China}

\author{Yang Jiang}
\email{corresponding author: jiangyang@itp.ac.cn}
\affiliation{CAS Key Laboratory of Theoretical Physics, 
	Institute of Theoretical Physics, Chinese Academy of Sciences,
	Beijing 100190, China}
\affiliation{School of Physical Sciences, 
	University of Chinese Academy of Sciences, 
	No. 19A Yuquan Road, Beijing 100049, China}

\author{Qing-Guo Huang}
\email{corresponding author: huangqg@itp.ac.cn}
\affiliation{CAS Key Laboratory of Theoretical Physics,
	Institute of Theoretical Physics, Chinese Academy of Sciences,
	Beijing 100190, China}
\affiliation{School of Physical Sciences,
	University of Chinese Academy of Sciences,
	No. 19A Yuquan Road, Beijing 100049, China}
\affiliation{School of Fundamental Physics and Mathematical Sciences
Hangzhou Institute for Advanced Study, UCAS, Hangzhou 310024, China}


\date{\today}
	
\begin{abstract}
Ultralight bosons are promising dark matter candidates and can trigger superradiant instabilities of spinning black holes (BHs), resulting in long-lived rotating ``bosonic clouds'' around the BHs and dissipating their energy through the emission of monochromatic gravitational waves (GWs). We focus on the scalar bosons minimally coupled with both isolated stellar-origin BHs (SBH) and their binary merger remnants, and perform Bayesian data analysis to search for the stochastic GW background from all the unstable modes that can trigger the superradiant instabilities using the data of Advanced LIGO and Advanced Virgo's first three observing runs. We find no evidence for such signal, and hence rule out the scalar bosons within the mass range $[1.5, 15]\times10^{-13}$ eV, $[1.8, 8.1]\times10^{-13}$ eV and $[1.3, 17]\times10^{-13}$ eV at $95\%$ confidence level for isolated SBHs having a uniform dimensionless spin distribution in $[0,1]$, $[0,0.5]$ and $[0.5,1]$, respectively.


\end{abstract}
	
	
\maketitle
	
	
\section{Introduction}
The first detection of gravitational waves (GWs) from a binary black hole (BH) \cite{LIGOScientific:2016aoc} and a binary neutron star \cite{LIGOScientific:2017vwq} has marked the beginning of GW astronomy, opening a new window to test fundamental physics in strong field regions \cite{Barack:2018yly,Baibhav:2019rsa,LIGOScientific:2019fpa,Abbott:2020jks,Sathyaprakash:2019yqt,Barausse:2020rsu} as well as providing a further understanding of BH population \cite{LIGOScientific:2018mvr,Abbott:2020gyp,LIGOScientific:2021psn}.

Ultralight bosons predicted in various models beyond Standard Model  
\cite{Arvanitaki:2009fg,Essig:2013lka,Irastorza:2018dyq,Goodsell:2009xc,Jaeckel:2010ni,Graham:2015rva,Agrawal:2018vin} are important dark matter (DM) candidate, and using GWs to probe ultralight bosons have aroused much attention recently \cite{Brito:2017wnc,Brito:2017zvb,Tsukada:2018mbp,East:2018glu,Ikeda:2018nhb,Dergachev:2019wqa,Ng:2019jsx,Chen:2019fsq,Palomba:2019vxe,Brito:2020lup,Dergachev:2020fli,Zhu:2020tht,Tsukada:2020lgt,Ng:2020ruv,Yuan:2021ebu,Guo:2021xao,Kalogera:2021bya,LIGOScientific:2021jlr,Zhang:2021mks}. The mechanism responsible for such GW observations is the so-called superradiant instability \cite{Damour:1976kh,Zouros:1979iw,Detweiler:1980uk,Dolan,string_axiverse,Shlapentokh-Rothman:2013ysa,Pani:2012vp,Pani:2012bp,Witek:2012tr,sr_tensor,Endlich:2016jgc,East:2017mrj,East:2017ovw,sr_vector_4,Cardoso:2018tly,East:2018glu,Frolov:2018ezx,Dolan:2018dqv,Baumann:2019eav,Brito:2020lup,Tsukada:2020lgt,Yuan:2021ebu}. Bosons with masses $m_s$ could form a quasi bound state around a rotating BH at a typical oscillation frequency $\omega_{R}\equiv m_s c^2/\hbar$ whenever the superradiant condition $0<\omega_R<m\Omega_{\mathrm{H}}$ is satisfied \cite{Press:1972zz,Detweiler:1980uk,Cardoso:2004nk,Dolan:2007mj,Brito:2015oca}. Here $c$ is the speed of light, $m$ is the azimuthal index and $\Omega_\mathrm{H}$ is the horizon angular velocity of the BH. The boson-BH system is unstable because the bosonic field would extract energy and angular momentum to undergo an exponential growth until reaching the saturation point where $\omega_R\sim m \Omega_\mathrm{H}$. This finally leads to a corotating and non-axisymmetric ``boson cloud'' around the BH which subsequently dissipates its energy due to the emission of GWs with frequency $f_0 = \omega_R /\pi $ \cite{Arvanitaki:2014wva,Arvanitaki:2016qwi,Baryakhtar:2017ngi,Brito:2017wnc,Brito:2017zvb,Isi:2018pzk,Ghosh:2018gaw,Palomba:2019vxe,Sun:2019mqb,Zhu:2020tht,Brito:2020lup,Ng:2020jqd}. For bosons with masses $\sim 10^{-12.5}$ eV, the typical frequency of such GW falls in the frequency band of current ground-based GW detectors and a large number of such GW sources in the Universe can result in a stochastic GW background (SGWB).

Despite various direct searches for the nearly monochromatic GWs produced by the boson clouds \cite{Palomba:2019vxe,Sun:2019mqb,Ng:2020ruv,Tsukada:2020lgt,LIGOScientific:2021jlr}, no such signals have been detected so far. The SGWB from boson clouds was calculated by \cite{Brito:2017zvb,Brito:2017wnc} and \cite{Tsukada:2018mbp} performed the first search of such SGWB in the first observing run of LIGO. The authors of \cite{Tsukada:2018mbp} excluded scalar bosons with masses $[2.0,3.8]\times10^{-13}$ eV under an optimistic assumption that the spin of isolated stellar origin BHs follows a uniform distribution. Recently, \cite{Tsukada:2020lgt} searched for the SGWB produced by vector bosons and excluded the mass range $[0.8,6.0]\times10^{-13}$ eV by assuming isolated BH population has a uniform distribution in $[0,1]$. However, these constraints on ultralight bosons only focus on the most unstable dipolar and quadrupolar modes, rendering an underestimate of the final SGWB and the corresponding mass range of the bosons \cite{Yuan:2021ebu}. 

In this work, we focus on a scalar bosonic field minimally coupled with both isolated stellar-origin BHs and their binary merger remnants and consider all higher unstable modes $m>1$ that would contribute to the SGWB. Then we perform Bayesian data analysis to search for ultralight scalar bosons using the data of LIGO-Virgo's first three observing runs \cite{KAGRA:2021kbb}. Throughout this work, we consider a $\Lambda$CDM model with $\Omega_{m}=0.3$, $\Omega_{\Lambda}=0.7$ and $H_0=67.4 \mathrm{km /s/Mpc}$ and from now on we use the units $G=c=1$.

\section{Superradiant instability and SGWB from boson clouds}
We focus on a real massive scalar field minimally coupled to gravity. For a Kerr BH with initial mass $M_i$ and initial dimensionless angular momentum $\chi_i=J_i/M_i^2$, the scalar field can have quasi-bound states with eigen-frequencies (see e.g.~\cite{Brito:2015oca}):
\e
\omega_{nlm}\equiv \omega_R+i\omega_I.
\q
Here  $\omega_R,\ \omega_I$ are real numbers, $n$ denotes the principal quantum number and $l$ indicates the quantum index related to the field angular momentum. For unstable modes $\omega_I>0$, the BH continuously transfers its mass and angular momentum to the scalar field until reaching the saturation point $\omega_R\sim m\Omega_H$ in a characteristic timescale, $\tau_{\mathrm{inst}}\equiv 1/\omega_I$. For modes with $m<l$, it takes a much longer time for the scalar field to grow than the modes with $l=m$. Hence it is safe to consider only the $l=m$ modes during the evolution of a boson-BH system.
Let $M_f$ and $J_f$ be the final mass and final angular momentum of the BH after the scalar field grows to its maximum mass $M_S^{\mathrm{max}}$ under the mode $m$ and we have \cite{Tsukada:2018mbp}
\e\label{Mfinal}
M_{f}=\frac{m^{3}-\sqrt{m^{6}-16 m^{2} \omega_{R}^{2}\left(m M_{i}-\omega_{R} J_{i}\right)^{2}}}{8 \omega_{R}^{2}\left(m M_{i}-\omega_{R} J_{i}\right)}.
\q
Due to the conservation of energy and angular momentum, $J_f$ can be written by 
\e
J_f = J_i -\frac{m}{\omega_R}(M_i-M_f).
\q
After the scalar field grow to its maximum mass, it dissipates its energy in a typical timescale $\taugw$, namely
\e
\tau_{\mathrm{GW}} = M_{f}\left(\frac{d \tilde{E}}{d t} \frac{M_{S}^{\max }}{M_{f}}\right)^{-1},
\q
where $\frac{d \tilde{E}}{d t}$ is the reduced GW flux \cite{Yoshino:2013ofa,Brito:2017zvb,Siemonsen:2019ebd} and we adopt the analytical results obtained in \cite{Yoshino:2013ofa} (for $l=m$ modes):
\e\label{Edot}
	\frac{d \tilde{E}}{d t}\approx\frac
	{
		16^{l+1} l(2 l-1) \Gamma(2 l-1)^{2} \Gamma(l+n+1)^{2}
		(\mu M)^{4l+10}
	}
	{
	n^{4 l+8}(l+1) \Gamma(l+1)^{4} \Gamma(4 l+3) \Gamma(n-l)^{2}
	},
\q
with $\Gamma$ being the Gamma function. This result provides a good approximation when $M\mu\ll l$ where $\mu\equiv m_s/\hbar$, and we have checked that the difference between numeric results and Eq.~(\ref{Edot}) is $\lesssim\mathcal{O}(10^{-7})$ when calculating the final SGWB.

Typically, we have $\taugw \gg \tauinst \gg M$, and hence it is safe to describe the boson-BH system using a quasi-adiabatic approximation  \cite{Brito:2014wla} such that the GW emission takes place only after the boson cloud reaches its maximum mass.
The total GW energy of a boson-BH system emitted in a duration time $\Delta t$ can then be evaluated by, \cite{Brito:2017zvb}, 
\e\label{eq:energy}
E_{\mathrm{GW}}=\frac{M_{S}^{\max } \Delta t}{\Delta t+\tau_{\mathrm{GW}}},
\q
where the signal duration time is $\Delta t=t_0-t_f$ \cite{Tsukada:2020lgt}, where $t_0$ is the age of the Universe and $t_f$ represents the time when the BH is formed.

Until now, our discussion is concentrated on a certain unstable mode $m$. However, all the unstable modes whose instability timescale $\tau_{\mathrm{inst}}$ shorter than the age of the Universe (or the age of a BH) will follow the above process and contribute to the final SGWB. In this work, we follow the hierarchical method proposed in \cite{Yuan:2021ebu} to include the contribution of higher modes. The mechanism responsible for this method is that $\tau_{\mathrm{inst}} \propto {(\mu M)}^{-(4l+4)}$ \cite{Detweiler:1980uk} when $\mu M \ll 1$, indicating that higher modes with increasing $l$ grow much slower than lower modes. As a result, it is safe to include all unstable modes starting from $l=m=1$ one by one using the description from Eq.~(\ref{Mfinal}) to Eq.~(\ref{eq:energy}). Similar to \cite{Yuan:2021ebu}, we only consider modes with $n=l+1$ since larger $n$ modes are quickly re-absorbed by the BH and have negligible contribution to the final SGWB \cite{Siemonsen:2019ebd}. 

\begin{figure}[ht]
    \centering
    \includegraphics[width=0.9\columnwidth]{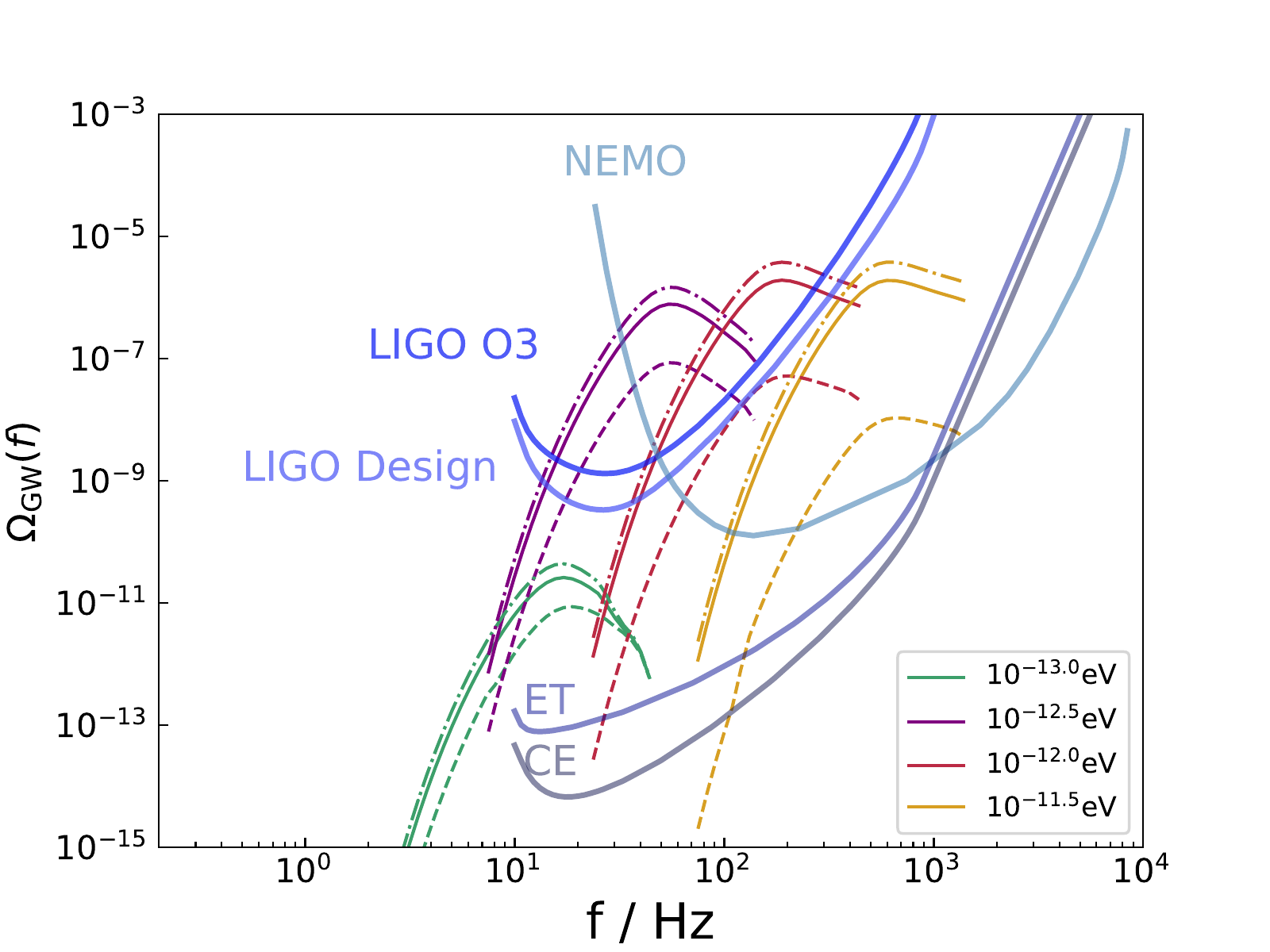}
    \caption{The SGWB from scalar clouds assuming uniform spin distribution for isolated SBHs. The results here include all the $m$ modes that contribute to the final SGWB. Dotted dashed lines, solid lines, and dashed lines correspond to $\chi_{i} \in [0.5,1]$, $\chi_{i} \in [0,1]$ and $\chi_{i} \in [0,0.5]$, respectively. The power-law integrated sensitivity curves \cite{Thrane:2013oya} for LIGO, Neutron Star Extreme Matter Observatory (NEMO) \cite{Ackley:2020atn}, Cosmic Explorer (CE) \cite{Evans:2016mbw} and the Einstein Telescope (ET) \cite{Punturo:2010zz,Maggiore:2019uih} are also shown, assuming a unity threshold signal-to-noise ratio and a four-year-detection. For NEMO/CE/ET, we assume two co-aligned, co-located and identical detectors.}
    \label{waveform}
\end{figure}

SGWB is the incoherent superposition of unresolvable GWs. The spectrum of SGWB can be described in terms of the dimensionless energy density parameter (see e.g.~\cite{Maggiore:1900zz}), defined as the GW energy density per logarithm frequency normalized by $\rho_c$, the critical energy needed for a spatially flat Universe:
\m
\Omega_{\mathrm{GW}}(f)&&\equiv {1\over \rho_c}\frac{d\rho_{\mathrm{GW}}}{d \ln(f)}.
\n
In this work, we consider boson clouds formed around both isolated stellar-origin BHs in extra-galaxy and BH remnants formed from binary BH mergers, namely
\e
\Omega_{\mathrm{GW}}(f)=\Omega^{iso}_{\mathrm{GW}}(f)+\Omega^{rem}_{\mathrm{GW}}(f). 
\q
By summing over all the GW sources in the sky, these two quantities can be evaluated as
\m
\Omega^{iso}_{\mathrm{GW}}(f)&&=\frac{f}{\rho_{c}} \int d \chi d M d z \frac{d t_L}{d z}  \frac{d E_{s}}{d f_{s}} \frac{d^{2} \dot{n}}{d M d \chi},\\
\Omega^{rem}_{\mathrm{GW}}(f)&&=\frac{f}{\rho_{c}} \int  d M_1 dM_2 d z \frac{d t_L}{d z}  \frac{d E_{s}}{d f_{s}} \mathcal{R}(z,M_1,M_2),~~
\n
where $d t_L/dz$ is the derivative of the lookback time with respect to the redshift $z$, ${d^{2} \dot{n}}/{d M d \chi}$ describes the fomation rate of SBHs per mass and per dimensionless spin and $\mathcal{R}(z,M_1,M_2)$ is the merger rate density for a binary BH with mass $M_1$ and $M_2$. 
For the BH population, we follow \cite{Yuan:2021ebu} and the references therein to obtain ${d^{2} \dot{n}}/{d M d \chi}$ and $\mathcal{R}(z,M_1,M_2)$. 
Since the BH poppulation model for isolated SBHs does not include spin distribution, we follow \cite{Brito:2017wnc,Brito:2017zvb} and assume uniform spin distribution for isolated SBHs. We focus on three scenarios, namely $\chi_i \in [0,1]$, $\chi_i \in [0,0.5]$ and $\chi_i \in [0.5,1]$ corresponding to neutral, less optimistic and optimistic scenarios, respectively. For SBH remnants, as suggested by Numerical Relativity simulations, the mass and dimensionless spin of a BH remnant formed from a non-spinning binary BH with masses $M_1$ and $M_2$ can be well estimated by \cite{Berti:2007fi,Scheel:2008rj,Barausse:2009uz}
\begin{eqnarray}\label{massspin}
M_i&=&M_1+M_2-(M_1+M_2)\Bigg[\left(1-\sqrt{\frac{8}{9}} \right)\nu\nonumber\\
&\qquad&-4\nu^2\left(0.19308+\sqrt{\frac{8}{9}} -1 \right)\Bigg],\nonumber\\
\chi_i&=&\nu (2\sqrt{3}-3.5171\nu+2.5763\nu^2),
\end{eqnarray}
where $\nu \equiv M_1M_2/(M_1+M_2)^2$ denotes the symmetric mass ratio. Note that we assume the binary BHs to be non-spinning to obtain the most conservative spin for the BH remnant, as well as the final SGWB. In addition, since the GWs emitted from boson clouds are nearly monochromatic, the energy spectrum of a single GW event in the source frame, ${d E_{s}}/{d f_{s}}$, can be approximated as \cite{Brito:2017wnc,Tsukada:2018mbp,Tsukada:2020lgt},
\e
\frac{d E_{s}}{d f_{s}} = E_{\mathrm{GW}} \delta\left(f(1+z)-f_{0}\right).
\q
The spectrum of SGWB from boson clouds is illustarted in Fig.~\ref{waveform}. Readers interested in the difference of the SGWB waveform between considering all higher $m$-modes and only considering the fundamental mode can refer to Fig.~2 in \cite{Yuan:2021ebu}.


\section{Data analysis and Results}
\begin{figure}[ht]
    \centering
    \includegraphics[width=0.9\columnwidth]{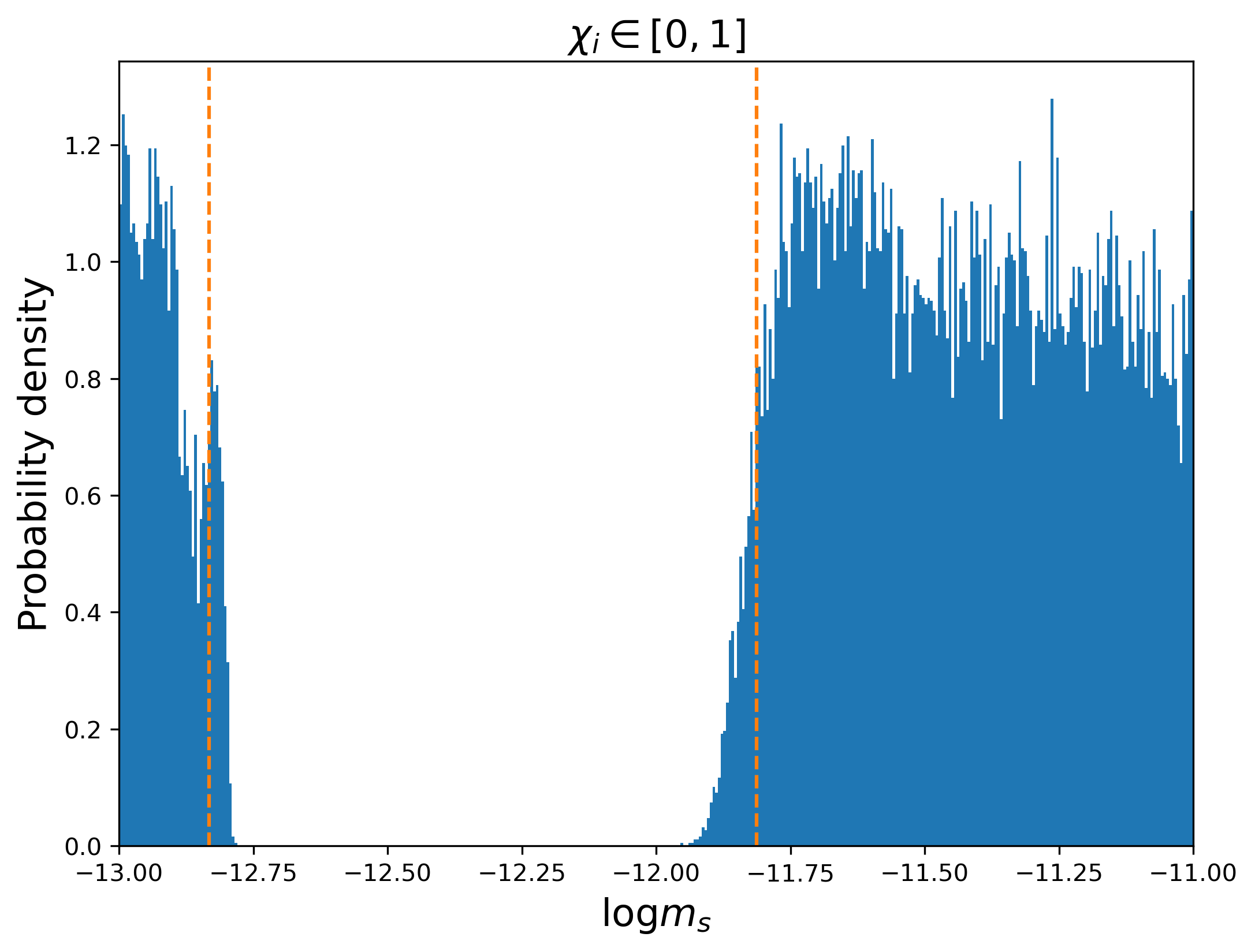}\\
    \includegraphics[width=0.9\columnwidth]{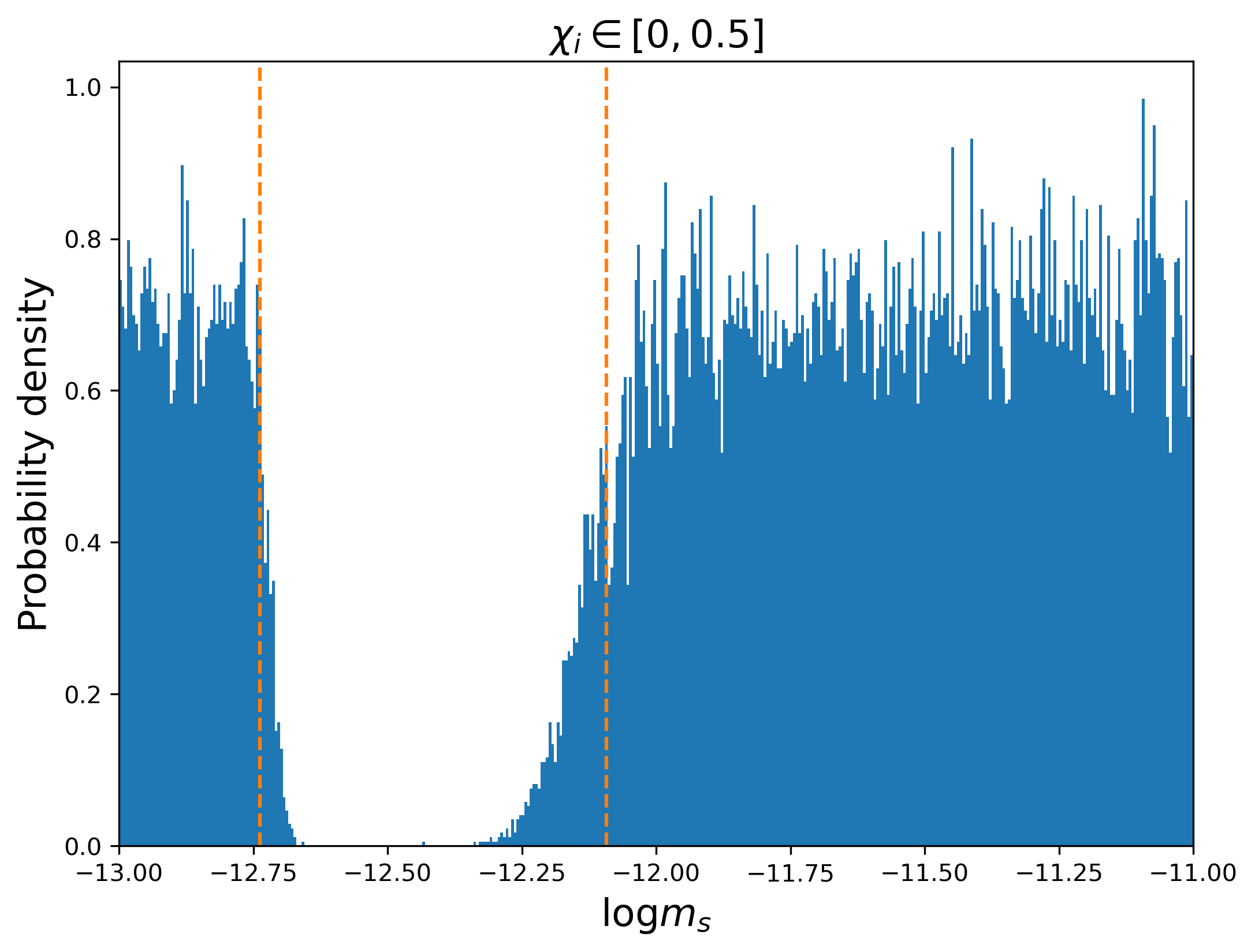}\\
    \includegraphics[width=0.9\columnwidth]{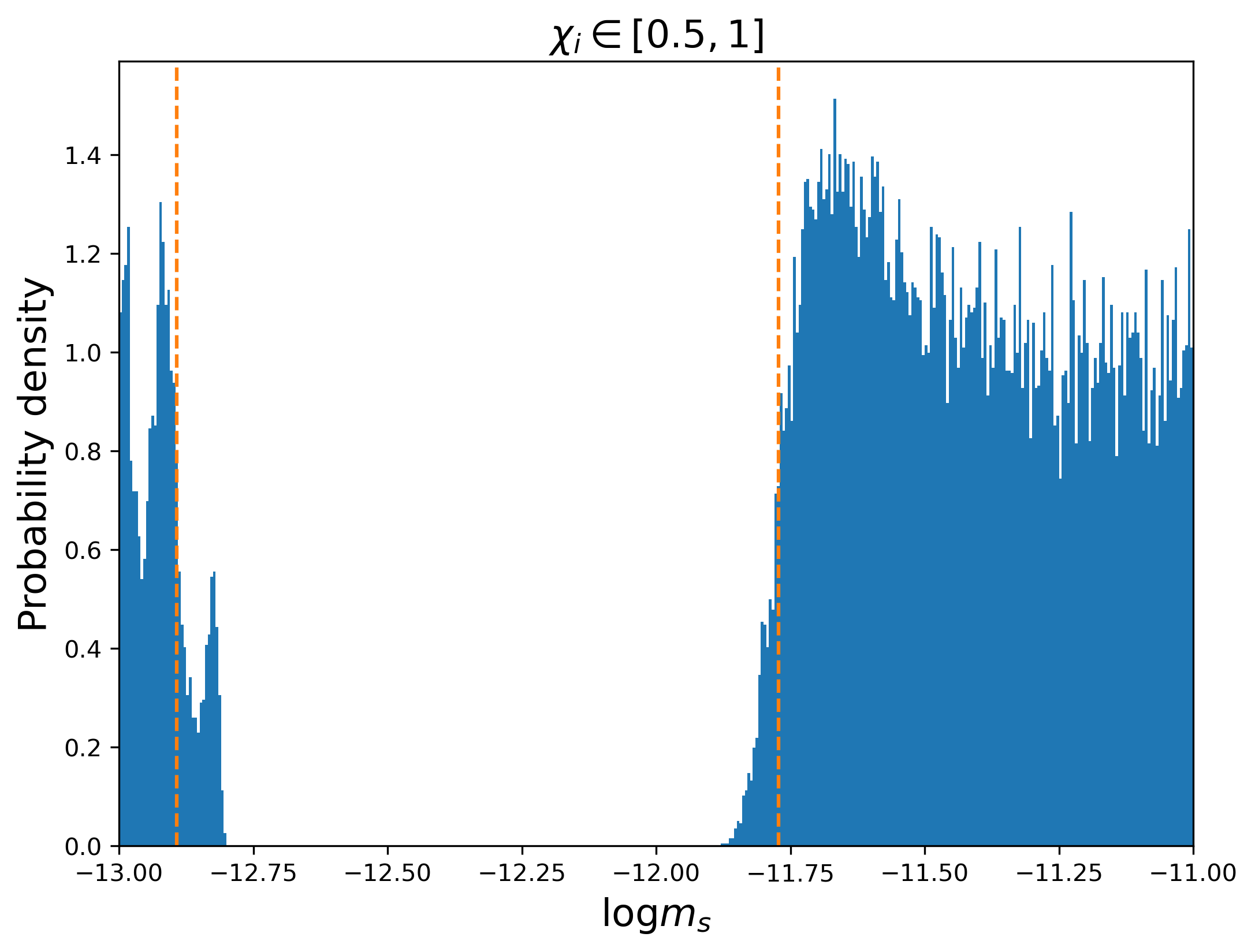}
    \caption{Posterior distributions of $\log m_s$ assuming a uniform spin distribution for the isolated stellar origin BHs. The orange dash lines denote $95\%$ exclusion intervals.}
    \label{fig:posterior}
\end{figure}
The detection of SGWB depends on coupling the strain data of two detectors and an estimator of $\Omega_\text{GW}(f)$ can be defined as \cite{Romano:2016dpx,KAGRA:2021kbb}
\begin{equation}
    \hat{C}_{IJ}(f)=\frac{2}{T}\frac{\text{Re}[\tilde{s}_I^*(f)\tilde{s}_J(f)]}{\gamma_{IJ}(f)S_0(f)}, 
\end{equation}
where $S_0(f)=(3H_0^2)/(10\pi^2f^3)$. $\gamma_{IJ}(f)$ is the overlap reduction function \cite{PhysRevD.59.102001} for the detector pair $IJ$ and $T$ denotes the observing time. We follow \cite{PhysRevLett.109.171102} and perform Bayesian inference to determine the mass of ultralight scalar bosons. The estimator $\hat{C}(f)$ of the SGWB energy spectrum has been calculated based on the observing data from Advanced LIGO's and Advanced Virgo's O1-O3 runs \cite{KAGRA:2021kbb}. Assuming Gaussian-distributed noise signal and the likelihood is given by
\begin{equation}
     p(\hat{C}_{IJ}|\bm{\theta};\lambda)\propto \exp\left[-\sum_{f}\frac{\left(\hat{C}_{IJ}(f)-\lambda\Omega(f;\bm{\theta})\right)^2}{2\sigma^2_{IJ}(f)}\right].
\end{equation}
Here $\bm{\theta}$ denotes the parameters to be determined and $\lambda$ is the calibration uncertainty of the detectors which is marginalized \cite{Whelan_2014}. To obtain a joint likelihood of overall detector pairs, we just need to multiply the likelihoods of every baseline $IJ$. According to the Bayes theorem, the posterior distribution is
\begin{equation}
    p(\bm{\theta},\mathcal{M}|\hat{C})\propto p(\hat{C}|\bm{\theta})p(\bm{\theta},\mathcal{M}).
\end{equation}
Evidence of the model $\mathcal{M}$ is calculated by marginalizing the parameters $\bm{\theta}$ appeared. The ratio of evidence
\begin{equation}
    \mathcal{B}_{12}=\frac{p(\hat{C}|\mathcal{M}_1)}{p(\hat{C}|\mathcal{M}_2)},
\end{equation}
also called the Bayes factor, is a criterion of preference between two different models. In this work, the Bayes factor between the signal-contained model and a pure noise model is used to show whether the signal exists or not.
We also perform a Bayesian analyse to recover an injected signal with $m_s =10^{-12.5}\,\mathrm{eV}$ and we obtain a Bayes factor $\log\mathcal{B}\simeq500$. Therefore, we conclude that our code can recover the injected signal successfully.

We set a log-uniform prior from $10^{-13}$ eV to $10^{-11}$ eV for the mass of scalar bosons $m_s$. The energy spectra of SGWB calculated under three different spin distribution are analyzed separately. Python package \texttt{bilby} \cite{Ashton_2019} is used to fulfill the algorithm above.

We find no such signals in the correlation spectrum during Advanced LIGO's and Advanced Virgo's observing runs since the log Bayes factors $-0.27$, $-0.15$ and $-0.30$ are too small to claim a detection. This is quite as expected that the frequency of SGWB is not in the sensitive band of detectors if the mass of the boson is too large or too small.
On the other hand, the intensity of SGWB is large enough to be excluded by the observing data for medium mass bosons. Therefore, we calculate exclusion intervals based on the posterior distributions illustrated in \Fig{fig:posterior}. We find $m_s$ with mass ranges of $[1.5, 15]\times10^{-13}$ eV, $[1.8, 8.1]\times10^{-13}$ eV and $[1.3, 17]\times10^{-13}$ eV are excluded for the neutral, less optimistic and optimistic scenarios, respectively. Details about the results are shown in \Table{tab:result}.

\begin{table}[ht]
    \centering
    \begin{tabular}{cccccc}
        \Xhline{0.09em}
        \multirowcell{2}{$\chi_i$\\(Uniform)} & \multicolumn{2}{c}{$m=1$} && \multicolumn{2}{c}{all $m$-modes} \\
        \Xcline{2-3}{0.01em}\Xcline{5-6}{0.01em}
         & $\log\mathcal{B}$ & $m_s$ (eV) && $\log\mathcal{B}$ & $m_s$ (eV) \\
        \hline
        $[0,1]$ & $-0.26$ & $[1.4,13]\times10^{-13}$ && $-0.27$ & $[1.5,15]\times10^{-13}$ \\
        $[0,0.5]$ & $-0.15$ & $[1.9,8.1]\times10^{-13}$ && $-0.15$ & $[1.8,8.1]\times10^{-13}$ \\
        $[0.5,1]$ & $-0.29$ & $[1.3,14]\times10^{-13}$ && $-0.30$ & $[1.3,17]\times10^{-13}$ \\
        \Xhline{0.09em}
    \end{tabular}
    \caption{Results of Bayesian inference and exclusion intervals for the mass of boson at $95\%$ credible level. The results in the second column only consider the most unstable mode $m=1$, while the results in the third column consider all the $m-$modes that contribute to the SGWB.  }
    \label{tab:result}
\end{table}

\section{Conclusion and Discussion}
In this work, we search for the SGWB signal produced by scalar boson clouds around both isolated SBHs and SBH remnants in the Advanced LIGO and Advanced Virgo's first three observing runs. We consider all the unstable modes of the bosonic field that would contribute to the final SGWB. No such signals are found and then we place constraints on the mass of scalar bosons. Assuming uniform spin distribution for isolated SBHs, we find that the scalar bosons in the mass range  $[1.5,15]\times10^{-13}$ eV, $[1.8,8.1]\times10^{-13}$ eV and $[1.3,17]\times10^{-13}$ eV are excluded for $\chi_i\in[0,1]$, $\chi_i\in[0,0.5]$ and $\chi_i\in[0.5,1]$,  respectively. 
The exclusion mass range is 
$[2.0,3.8]\times10^{-13}$ eV from LIGO O1 data by only considering the $m=1$ mode and assuming a uniform spin distribution for the isolated SBHs in \cite{Tsukada:2018mbp}. The improvements on the limits of the exclusion mass range in this paper mainly come from the better sensitivity of O3 than O1 and including all the $m$ modes that contribute to the SGWB.


\section*{Acknowledgments}
We acknowledge the use of \texttt{GWSC.jl} package \cite{GWSC} for plotting the sensitivity curves. CY wishes to thank Yu-jia Zhai for her daily encouragement. This work is supported by the National Key Research and Development Program of China Grant No.2020YFC2201502, grants from NSFC (grant No. 11975019, 11991052, 12047503), Key Research Program of Frontier Sciences, CAS, Grant NO. ZDBS-LY-7009, CAS Project for Young Scientists in Basic Research YSBR-006, the Key Research Program of the Chinese Academy of Sciences (Grant NO. XDPB15).

\bibliography{./ref}
	
\end{document}